\documentclass[journal]{IEEEtran}
\usepackage{amsmath,amsfonts,bm,amssymb,psfrag,ifthen,color,subfigure}
\usepackage{mathrsfs}
\usepackage[justification=centering,font=footnotesize]{caption}
\usepackage{todonotes}
\usepackage[bottom]{footmisc}

\allowdisplaybreaks[4]

\usepackage[dvips]{epsfig}
\usepackage[dvips]{graphicx}
\usepackage{amsmath,amsfonts,bm,amssymb,psfrag,ifthen,color,subfigure}
\usepackage{algpseudocode}
\usepackage{algorithm}
\usepackage{algorithmicx}
\usepackage{ifthen}
\usepackage{setspace}
\usepackage{cite}
\usepackage{url}
\usepackage{longtable}
\usepackage{stfloats}
\usepackage{graphics,booktabs,color,subfigure}
\usepackage{float}
\usepackage{diagbox}
\usepackage{threeparttable}
\usepackage{tabularx}
\usepackage{hyperref} % 超链接
\usepackage{caption}
\usepackage{amsmath}
\usepackage{mathabx}
\usepackage{footnote}
\usepackage{caption}
\newcommand{\bd}{\boldsymbol}

\usepackage{listings}

\floatname{algorithm}{Algorithm}

\begin{document}
\title{Joint Beamforming and Illumination Pattern Design for Beam-Hopping LEO Satellite Communications}
\author{Jing Wang,~\IEEEmembership{Student Member,~IEEE,} Chenhao Qi,~\IEEEmembership{Senior Member,~IEEE,} \\Shui~Yu,~\IEEEmembership{Fellow,~IEEE,} and Shiwen~Mao,~\IEEEmembership{Fellow,~IEEE}
\thanks{Jing Wang and Chenhao Qi are with the School of Information Science and Engineering, Southeast University, Nanjing, 210096, China. (e-mail:~\{wangjing12,~qch\}@seu.edu.cn).} 
\thanks{Shui~Yu is with the Faculty of Engineering and Information Technology, University of Technology Sydney, Australia. (e-mail:~ Shui.Yu@uts.edu.au).}
\thanks{Shiwen~Mao is with the Department of Electrical and Computer Engineering, Auburn University, USA. (e-mail:~ smao@ieee.org).}
}

\markboth{Accepted by IEEE Transactions on Wireless Communications}
{}

\maketitle
\IEEEpeerreviewmaketitle

\begin{abstract}
Since hybrid beamforming (HBF) can approach the performance of fully-digital beamforming (FDBF) with much lower hardware complexity, we investigate the HBF design for beam-hopping (BH) low earth orbit (LEO) satellite communications (SatComs). Aiming at maximizing the sum-rate of totally illuminated beam positions during the whole BH period, we consider joint beamforming and illumination pattern design subject to the HBF constraints and sum-rate requirements. To address the non-convexity of the HBF constraints, we temporarily replace the HBF constraints with the FDBF constraints. Then we propose an FDBF and illumination pattern random search (FDBF-IPRS) scheme to optimize illumination patterns and fully-digital beamformers using constrained random search and fractional programming methods. To further reduce the computational complexity, we propose an FDBF and illumination pattern alternating optimization (FDBF-IPAO) scheme, where we relax the integer illumination pattern to continuous variables and after finishing all the iterations we quantize the continuous variables into integer ones. Based on the fully-digital beamformers designed by the FDBF-IPRS or FDBF-IPAO scheme, we propose an HBF alternating minimization algorithm to design the hybrid beamformers. Simulation results show that the proposed schemes can achieve satisfactory sum-rate performance for BH LEO SatComs.

\end{abstract}
%\newpage
\begin{IEEEkeywords}
Beam-hopping (BH), hybrid beamforming (HBF), illumination pattern, LEO satellite communications.
\end{IEEEkeywords}

\section{Introduction}
To achieve full coverage of spatial and terrestrial wireless communications, the upcoming sixth generation wireless communications integrating satellite communications (SatComs) will establish a fully connected wireless network. SatComs can provide seamless and stable wireless service to complement and extend terrestrial communications, and therefore can address the challenge of insufficient connectivity for remote areas such as deserts, mountains and oceans~\cite{Oltjon_COMST_2021,Qi2024KeyIssues,Xiao_WC_2024}. Compared to medium earth orbit and geosynchronous earth orbit counterparts, the low earth orbit (LEO) satellites are proximal to the earth, with the advantages of low latency in wireless access, reduced energy for launching, and small power for signal transmission from the satellites to terrestrial receivers~\cite{Xing_ICM_2021}. Therefore, LEO SatComs have received increasing attention and become hotspots of commercial investment. The companies such as SpaceX and Amazon, have already put forward their commercial LEO SatCom products including Starlink and Kuiper~\cite{Portillo_Acta_2019}.

As one of the key technologies of LEO SatComs, beam-hopping (BH) has raised great interest from both academia and the industry, owing to its flexibility and low complexity for implementation~\cite{Xing_ICM_2021,Lyu_CC_2023}. BH is capable of achieving multi-beam coverage with reduced inter-beam interference~\cite{Ha_2022_GLOBECOM,Lin_TWC_2022} and improved resource utilization~\cite{Anyue2022TWC,Li2023TCCN}, by simultaneously activating a set of beams at each time slot in a designed illumination pattern and periodically repeating in the next BH time window.  To reduce the inter-beam interference among illuminated beams in the same time slot, adjacent beams are not preferred by the BH~\cite{Ha_2022_GLOBECOM}. Aiming at minimizing the interference-based penalty function, a dynamic BH scheme is proposed in conjunction with selective precoding~\cite{Lin_TWC_2022}. To appropriately allocate the resource in SatComs, a joint power allocation and  BH design scheme with non-orthogonal multiple access is employed subject to the requested traffic demands~\cite{Anyue2022TWC}. Aiming at efficiently utilizing time domain and power domain resources in BH SatComs, an adaptive resource adjustment method is proposed to design the BH illumination pattern and transmit power allocation~\cite{Li2023TCCN}.

%Furthermore, beamforming has been widely adopted to achieve high data-rate for LEO SatComs. Compared to the fully-digital beamforming (FDBF) that achieves the satisfactory performance at the cost of very-high hardware complexity~\cite{Zhou_TVT_2022,Qi2022Energy}, hybrid beamforming (HBF) is promising to balance the performance and hardware constraints, especially for the LEO SatComs with limited on-board resource~\cite{You_JSAC_2020}. To maximize the energy efficiency, a hybrid analog and digital precoding designing method is developed in the massive MIMO LEO SatCom systems~\cite{Li_TWC_2022}, and an HBF scheme with resource allocation is proposed subject to transmit power limitation~\cite{Peng_TVT_2021}. Based on the angel information instead of channel state information in SatComs, an HBF scheme is designed using the codebook method to perform the beam selection for the LEO satellite~\cite{Shi2023TOB}.
Furthermore, the beamforming has been widely adopted to achieve high data-rate for LEO SatComs. Aiming at maximizing the sum-rate of the SatCom system, a joint beamforming and power allocation scheme is proposed to mitigate the interference among the ground users~\cite{Gao2023TOC}. Compared to the fully-digital beamforming (FDBF) that achieves the satisfactory performance at the cost of very-high hardware complexity~\cite{Zhou_TVT_2022,Qi2022Energy}, hybrid beamforming (HBF) is promising to balance the performance and hardware constraints. A hybrid analog and digital beamforming method is developed in the massive MIMO LEO SatCom systems~\cite{Li_TWC_2022}, where the HBF architecture is equipped at the satellite to transmit multiple beams. To further mitigate intra-beam and inter-beam interference, a two-stage HBF scheme together with an adaptive user scheduling scheme is proposed for millimeter-wave spectrum coexisting integrated terrestrial-satellite network~\cite{Peng_TVT_2021}. Based on the angel information instead of channel state information in SatComs, an HBF scheme is designed using the codebook to perform the beam selection for the LEO SatComs~\cite{Shi2023TOB}. To achieve flexible beam coverage while maintaining the RF cost for the LEO SatComs, the beamforming based on movable antenna array is considered with the time-varying beam coverage for terrestrial users~\cite{Zhu_CMag_2024},~\cite{Zhu_arXiv_2024}.

%Based on the correlation of the uplink-downlink channel state information (CSI) between the LEO satellites and ground users, a deep learning based downlink CSI prediction scheme is designed to auxiliary implement the HBF~\cite{Zhang2022IOTJ}.

By combining the advantages of the flexibility and high data-rate, the beamforming design is considered for SatComs with BH. To reduce the power consumption, compressed sensing is employed to design FDBF with BH~\cite{Ha_2022_GLOBECOM}. To balance the data traffic among clusters where each cluster is formed by multiple beams, a joint optimization method of singular-value-decomposition FDBF and BH is proposed subject to the service requirement of each beam~\cite{Chen2021CC}. Aiming at maximizing the energy efficiency in BH SatComs, a cluster-based BH scheme with FDBF design is proposed to achieve the on-demand capacity allocation~\cite{Yang2023Cluster}. However, to the best knowledge of authors, so far there has been no work on HBF for LEO SatComs with BH aiming at sum-rate maximization. Note that the sum-rate maximization is an important objective for wireless communications including SatComs and the HBF can effectively reduce the hardware complexity for the payload of the satellite.

In this paper, we investigate the HBF design for BH SatComs. Aiming at maximizing the sum-rate of totally illuminated beam positions during the whole BH period, we consider joint beamforming and illumination pattern (BIP) design subject to the HBF constraints and sum-rate requirements. As a summary, our contributions include the following three points, where the first point is included in our conference paper~\cite{JingGLOBECOM2023}.

\begin{itemize}
\item To address the non-convexity of the HBF constraints, we temporarily replace the HBF constraints with the FDBF
constraints. Then an FDBF and illumination pattern random search (FDBF-IPRS) scheme is proposed, where the illumination pattern of SatComs is generated by the constrained random search and then the fully-digital beamformers of the satellite are designed by the fractional programming (FP) methods. 

%We focus on sum-rate maximization of joint BIP design problem. To solve this joint BIP design problem, we propose an FDBF-HBF and illumination pattern random search (FDBF-HBF-IPRS) scheme as well as an FDBF-IPRS-HBF scheme. For the FDBF-HBF-IPRS scheme, a candidate illumination pattern set is generated by random search method. Then we achieve the optimized FDBF beamformers for each candidate illumination pattern using fractional programming (FP) method. Then we find the final FDBF beamformers achieving the largest sum-rate, where the designed FDBF beamformers are used to design HBF by minimizing the Euclidean distance between the designed FDBF beamformers and HBF beamformers.

\item To reduce the computational complexity, we propose an FDBF and illumination pattern alternating optimization (FDBF-IPAO) scheme to decouple the BIP problem into two subproblems, namely, the FDBF design subproblem and the illumination pattern design subproblem, and alternately optimize these two subproblems until a stop condition is triggered. For the FDBF design subproblem, the FP method is employed to obtain the fully-digital beamformers. For the illumination pattern design subproblem that is mixed-integer and non-convex, we relax the integer illumination pattern to continuous variables so that the FP method can be used for the optimization. Once the stop condition is triggered, we quantize the continuous variables into integer ones to determine the illumination pattern.

\item Based on the fully-digital beamformers designed by the FDBF-IPRS or FDBF-IPAO scheme, we propose an HBF alternating minimization (HBF-AM) algorithm to design the hybrid beamformers. By minimizing the Euclidean distance between the designed fully-digital beamformers and the hybrid beamformers, the digital beamformers and analog beamformers are iteratively optimized using the Riemannian manifold optimization method.

\end{itemize}

\emph{Notations}: Boldfaced lowercase and uppercase letters represent vectors and matrices, respectively. The conjugate, expectation, transpose, Hermitian transpose, Frobenius norm, and Kronecker product are denoted as ${\left(  \cdot  \right)^*}$, ${\mathbb{E}}{\left(  \cdot  \right)}$, ${\left(  \cdot  \right)^{\rm T}}$,  ${\left(  \cdot  \right)^{\rm H}}$, $\left\|  \cdot  \right\|_2$, and $\otimes$, respectively. $\mathbb{C}$ and $\mathbb{R}$ represent the sets of complex-valued numbers and real-valued numbers, respectively. $\bd I_{N_{\rm s}}$ denotes the Identity matrix with the dimension of $N_{\rm s} \times N_{\rm s}$. ${\cal M} \triangleq \left\{ {1,2,...,M} \right\}$, $\mathcal{N} \triangleq \left\{ {1,2,...,N_{\rm s}} \right\}$ and  ${\mathcal{N}}_{\rm BS} \triangleq \left\{ {1,2,...,N_{\rm BS}} \right\}$.

%-----------------------------------------------------------------------------------------------------	

\begin{figure}[!t]
	\centering
	\includegraphics[width=0.42\textwidth]{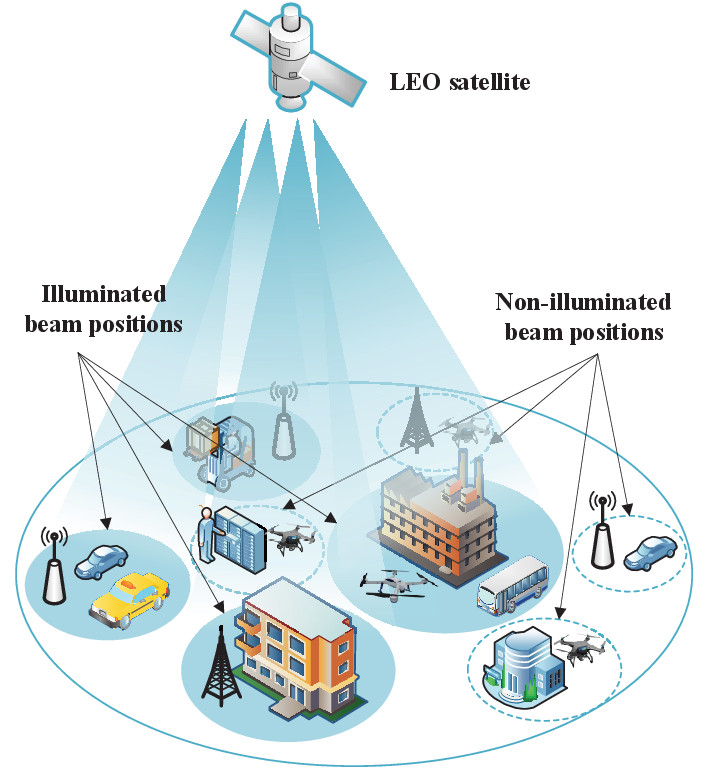}
	\caption{Illustration of system model.}
	\label{fig1}
\end{figure}

\section{System Model}
Consider a downlink LEO SatCom system, where the LEO satellite equipped with $N_{\rm BS}$ antennas can simultaneously generate $K$ spot beams to illuminate $K$ beam positions on the ground. To illuminate the total $N_{\rm s}$ beam positions, we use the BH, where each BH period includes $M$ time slots. To guarantee that each beam position is illuminated at least once during each BH period, we require 
\begin{equation}
	KM \geq N_{\rm s}. 
\end{equation}
As illustrated in Fig.~\ref{fig1}, the beam positions with different radius are determined according to the number of ground users and the service demand before performing the BH~\cite{Qi2020Multicast}. Note that the satellite works in the multicast mode and therefore different users located in the same beam position receive the same signal from the satellite.

To transmit $K$ independent data streams at each time slot,  the LEO satellite employs the HBF architecture equipped with $N_{\rm RF}$ RF chains, where $K \le N_{\rm RF}$. The HBF architecture includes analog beamformer ${{\bd{F}}_t} \in { \mathbb{C}^{N_{\rm BS} \times N_{\rm RF}}}$ and digital beamformer ${{\bd{Q}}_t} \in { \mathbb{C}^{N_{\rm RF} \times N_{\rm s}}}$ for the $t$th time slot, respectively. Let $\bd s \in { \mathbb{C}^{N_{\rm s}}}$ denote a symbol vector transmitted by the LEO satellite. The received signal at the total $N_{\rm s}$ beam positions in the $t$th time slot can be expressed as
\begin{align}\label{y_signal}
	{{\bd{y}}_t} = {\bd{H}}{{\bd{F}}_t}{{\bd{Q}}_t}{{\bd{s}}} + {\bd{z}},~ t\in {\cal M},
\end{align}
where $\bd z \in { \mathbb{C}^{N_{\rm s}}}$ denotes an additive white Gaussian noise vector with ${\left[ {\boldsymbol{z}} \right]_{n}} \sim{\mathcal {CN}}\left( {0,{\sigma ^2}} \right)$, and ${\boldsymbol{H}} \in {\mathbb{C}}^{N_{\rm s} \times N_{\rm BS}}$ denotes the downlink SatCom channel matrix. Note that ${\boldsymbol{H}}$ is supposed to be the same during the whole BH period~\cite{Lin_TWC_2022}.  In fact, we have 
	\begin{align}
		\bd{H} \triangleq[\boldsymbol{h}_1^{\rm T},\boldsymbol{h}_2^{\rm T},\ldots,\boldsymbol{h}_{N_{\rm s}}^{\rm T}]^{\rm T},
	\end{align}
	where $\bd{h}_n$ denotes the channel vector between the LEO satellite and the $n$th beam position for $n\in \mathcal{N}$. Assuming that the Doppler frequency shift caused by the satellite movement is compensated by the ground users and the sky is clear so that the rain attenuation is neglected, we can express $\boldsymbol{h}_n$ as
	\begin{align}\label{channel1}
		{{{\bd{h}}_n}} = \sum\nolimits_{l = 1}^{{L_n}} {{g_l^{\left(n\right)}}{\bd{v}}^{\rm H}\big( {{N_{{\rm{BS}}}},{\theta _l^{\left(n\right)}}} \big)} ,
	\end{align}
	where $L_n$ denotes the number of the channel paths, ${\theta _l^{\left(n\right)}}$ represents the angle-of-departure of the $l$th channel path, and ${\bd{v}}\big( {{N_{{\rm{BS}}}}, {\theta _l^{\left(n\right)}}} \big)$ denotes the channel steering vector of a uniform linear array given as
	\begin{align}
		{\bd{v}}\big( {{N_{{\rm{BS}}}},{\theta _l^{\left(n\right)}}} \big) = \frac{1}{{\sqrt {{N_{{\rm{BS}}}}} }}\big[ {1,{e^{ j\pi {\theta _l^{\left(n\right)}}}},...,{e^{  j\pi \left( {{N_{{\rm{BS}}}} - 1} \right){\theta _l^{\left(n\right)}}}}} \big]^{\rm T},
	\end{align}	
	for $l=1,2,...,{L_n}$. In fact, we have ${\theta _l^{\left(n\right)}} = 2d_0 \frac{f_{\rm c}}{v_{\rm c}}\sin{\tilde{\theta}_l^{\left(n\right)}}$, where $d_0$, $f_{\rm c}$ and $v_{\rm c}$ represent the antenna spacing, carrier frequency and speed of light, respectively, and $\tilde{\theta}_l^{\left(n\right)}$ denotes the physical angle of  the $l$th channel path for $l=1,2,...,{L_n}$. Note that $g_l^{\left(n\right)}$ represents the channel gain of the $l$th channel path obeying Rician fading distribution with Rician factor ${\chi  _l^{\left(n\right)}}$ and $\mathbb{E}\big( {{{\big| {{g_l^{\left(n\right)}}} \big|}^2}} \big) = {\eta _l^{\left(n\right)}}$. Specifically, ${\eta _l^{\left(n\right)}}$ can be expressed as
	\begin{align}
		{\eta _l^{\left(n\right)}} = {\left( {\frac{v_{\rm c}}{{4\pi f_{\rm c} d}}} \right)^2}\frac{{G_{\rm{r}}}{G_{\rm t}}{N_{\rm BS}}}{{\kappa B{T_{\rm{R}}}}},
	\end{align}
	where $B$, $d$, $\kappa$,  and $T_{\rm R}$ represent the bandwidth, propagation distance, Boltzmann's constant and receiving noise temperature, respectively. $G_{\rm{r}}$ and ${G_{\rm t}}$ denote the transmitting antenna gain of the satellite and receiving antenna gain of the ground users, respectively~\cite{Li_TWC_2022}.

We denote a binary variable  ${x_{n,t}} \in \left\{ {0,1} \right\}$ as the illumination indicator of the $n$th beam position in the $t$th time slot, for $t\in {\cal M}$ and $n\in {\cal N}$. If ${x_{n,t}}  = 1$, the $n$th beam position is illuminated by a spot beam of the LEO satellite at the $t$th time slot; otherwise, it is not illuminated. Since the LEO satellite can simultaneously generate at most $K$ spot beams in each time slot, we have
\begin{equation}
 \sum\nolimits_{n = 1}^{N_{\rm s}} {{x_{n,t}}}   \le K.
\end{equation}
We define the illumination pattern as $\bd{X}$, where
\begin{equation}
	[\boldsymbol{X}]_{n,t} \triangleq  {{x_{n,t}}},~n\in\mathcal{N},~t\in\mathcal{M}.
\end{equation}

To simplify the notation, we define 
\begin{equation}\label{DBFcolumns}
	{{\bd{Q}}_t} \triangleq [{\bd{q}}_1^t,{\bd{q}}_2^t, \ldots ,{\bd{q}}_{N_{\rm s}}^t].
\end{equation}
Then, the achievable rate of the $n$th beam position in the $t$th time slot can be expressed as
\begin{align}\label{rate2}
	{R_{n,t}}\left( {{{\bd{F}}_t},{{\bd{Q}}_t},\bd{X}} \right) = {\log _2}\left( {1 + \frac{{{x_{n,t}}{{\left| {{{\bd{h}}_n}{{\bd{F}}_t}{{\bd{q}}_n^t}} \right|}^2}}}{\sum\nolimits_{k \ne n}^{{N_{{\rm{s}}}}} {{x_{k,t}}{{\left| {{{\bd{h}}_n}{{\bd{F}}_t}{{\bd{q}}_k^t}} \right|}^2}}  + {\sigma ^2}}}  \right).
\end{align}

%-----------------------------------------------------------------------------------------------------
\section{Joint Beamforming and Illumination Pattern Design}
To maximize the sum-rate of the total $N_{\rm s}$ beam positions during the whole BH period, we jointly optimize the analog and digital beamforming matrices $\{\bd{F}_t,\bd{Q}_t,t\in\mathcal{M}\}$ together with the illumination pattern. 
The joint BIP design problem can be formulated as
\begin{subequations}\label{orig_pro}
	\begin{align}
		\max_{{\bd X},\left\{ { {{\bd{F}}_t},{{\bd{Q}}_t,t \in {\cal M}} } \right\}} &\sum\limits_{n = 1}^{N_{\rm s}} {\sum\limits_{t = 1}^M {{R_{n,t}}\left( {{{\bd{F}}_t},{{\bd{Q}}_t},\bd{X}} \right)} } \label{orig_pro_a} \\
        {\rm{s}}{\rm{.t}}{\rm{.}}~~~~&\sum\limits_{t = 1}^M {{R_{n,t}}\left( {{{\bd{F}}_t},{{\bd{Q}}_t},\bd{X}} \right)}  \ge {{\gamma}_n},\forall n \in {\cal N},\label{orig_pro_b}\\
        &\sum\limits_{n = 1}^{N_{\rm s}} {{{\left\| {{{\bd{F}}_t}{{\bd q}_n^t}} \right\|}_2^2}}  \le {P_{{\rm{tot}}}},\forall t \in {\cal M},\label{orig_pro_c}\\
        &\left| {[{{\bd{F}}_t}]_{i,n}} \right| = 1,\forall i \in {{\cal N}_{{\rm{BS}}}},n \in {{\cal N}},\label{orig_pro_d}\\
		&\sum\nolimits_{n = 1}^{N_{\rm s}} {{x_{n,t}}}  \le K,\forall t \in {\cal M},\label{orig_pro_e}\\
		&{x_{n,t}} \in \left\{ {0,1} \right\},\forall n\in {\cal N},\forall t \in {\cal M}, \label{orig_pro_f}
	\end{align}
\end{subequations}
where $\gamma_n$ and $P_{\rm tot}$ are the predefined threshold of the sum-rate and the power constraint, respectively. To be specific, constraint \eqref{orig_pro_b} indicates that the sum-rate of the $n$th beam position during the whole BH period is no smaller than $\gamma_n$. Constraint \eqref{orig_pro_c} indicates that the total transmit power of the LEO satellite in each time slot is no larger than ${P_{{\rm{tot}}}}$. Constraint \eqref{orig_pro_d} indicates the unit-modulus constraints of analog beamformers. Constraint \eqref{orig_pro_e} indicates that the maximum number of illuminated beam positions in each time slot is no larger than $K$. In fact, the constraints \eqref{orig_pro_c}, \eqref{orig_pro_d} and \eqref{orig_pro_e} compose the HBF constraints. 

It is seen from \eqref{orig_pro_a} and \eqref{orig_pro_b} that the beamforming matrices and the illumination pattern are coupled, leading \eqref{orig_pro} to be a mixed integer non-convex optimization problem. To solve this joint BIP design problem, we temporarily replace the HBF constraints with the FDBF constraints and then design the illumination pattern.

%To solve this joint BIP design problem, we temporarily replace the HBF constraints by the FDBF constraints and generate a set of candidate illumination patterns. Based on the candidate illumination pattern set, we solve the FDBF design problem and achieve the optimized FDBF beamformers, which are used in an Euclidean distance minimization problem to design the HBF and we find one achieving the largest sum-rate. To reduce the computational complexity, we firstly achieve the designed FDBF beamformers and illumination pattern by choosing the largest sum-rate, which are used to design HBF.  To reduce the difficulty in tackling integer variables, an HBF-IPAO scheme is proposed, where we relax the integer variables to continuous variables and then quantize the continuous variables into integer ones after finishing all the iterations. To further reduce the complexity, we propose an FDBF-IPAO scheme to alternately optimize the FDBF beamformers and the relaxed variables, where the designed illumination pattern is achieved by quantization and then we design HBF based on the designed FDBF beamformers and illumination pattern.

Define the fully-digital beamformer at the $t$th time slot for $t\in \mathcal{M} $ as
\begin{equation}
	{{\bd{P}}_t} \triangleq [{\bd{p}}_1^t,{\bd{p}}_2^t, \ldots ,{\bd{p}}_{N_{\rm s}}^t] \in { \mathbb{C}^{N_{\rm BS} \times N_{\rm s}}}.
\end{equation}
We temporarily replace the HBF constraints on $\bd{F}_t$ and $\bd{Q}_t$, with the FDBF constraints on $\bd{P}_t$. Then the fully-digital BIP design problem can be given as
\begin{subequations}\label{orig_pro1}
	\begin{align}
		\max_{\boldsymbol{X},\left\{ {{{\boldsymbol{P}}_t,t\in\mathcal{M}}} \right\}}& \sum\limits_{n = 1}^{N_{\rm s}} {\sum\limits_{t = 1}^M {{R_{n,t}}\left( {{{\bd{P}}_t},{\bd X}} \right)} }  \label{orig_pro1_a}\\
		{\rm{s}}{\rm{.t}}{\rm{.  }}~~~~&\sum\limits_{t = 1}^M {{R_{n,t}}\left( {{{\bd{P}}_t},{\bd X}} \right)}  \ge {{\gamma}_n},\forall n \in {\mathcal N},\label{orig_pro1_b}\\
		&\sum\limits_{n = 1}^{N_{\rm s}} {{{\left\| {{{\boldsymbol{p}}_n^t}} \right\|}_2^2}}  \le {P_{{\rm{tot}}}},\forall t \in {\cal M},\label{orig_pro1_c}\\
		&\eqref{orig_pro_e},\eqref{orig_pro_f},
	\end{align}
\end{subequations}
which is also a mixed integer non-convex optimization problem. 

To solve~\eqref{orig_pro1}, we propose the FDBF-IPRS and FDBF-IPAO schemes in Section~\ref{subsubsec.FDBF-IPRS} and Section~\ref{subsubsec.FDBF-IPAO}, respectively. For the FDBF-IPRS scheme, the illumination pattern of SatComs is generated by
the constrained random search and then the fully-digital beamformers of the satellite are designed by the FP methods. For the FDBF-IPAO scheme, we decouple the fully-digital BIP problem into two subproblems, namely, the FDBF design subproblem and the illumination pattern design subproblem, and alternately optimize these two subproblems until a stop condition is triggered. For the FDBF design subproblem, the FP method is employed to obtain the fully-digital beamformers. For the illumination pattern design subproblem that is mixed-integer and non-convex, we relax the integer illumination pattern to continuous variables so that the FP method can be used for the optimization. Once the stop condition is triggered, we quantize the continuous variables into integer ones to determine the illumination pattern. 

To solve~\eqref{orig_pro}, in Section~\ref{subsection.HBF-AM} we propose an HBF-AM algorithm to design the hybrid beamformers based on the fully-digital beamformers designed by the FDBF-IPRS or FDBF-IPAO scheme. By minimizing the Euclidean distance between the designed fully-digital beamformers and the hybrid beamformers, the digital beamformers and analog beamformers are iteratively optimized using the Riemannian manifold optimization method.

%-----------------------------------------------------------------------------------------------------
%\subsection{IPRS Schemes for Joint BIP Problem}\label{subsec.FDBF}

\subsection{FDBF-IPRS Scheme}\label{subsubsec.FDBF-IPRS}
Suppose the candidate illumination patterns are denoted by a set $\left\{{\cal X}_i,i \in {\cal I}\right\}$, where $I$ is the number of candidate illumination pattern sets and ${\cal I} \triangleq \left\{ {1,2,...,I} \right\}$. To reduce the complexity of exhaustive search, we generate $\left\{{\cal X}_i, i \in {\cal I}\right\}$ obeying the following two criteria. 
\begin{itemize}
	\item To completely utilize the resource in each time slot, we simultaneously illuminated $K$ beams in the $t$th time slot, i.e., replacing the inequality by the equality in \eqref{orig_pro_e} as
 \begin{align}
	\sum\limits_{n = 1}^{N_{\rm s}} {{x_{n,t}}}  = K,~\forall t \in {\cal M}.\label{CritriaIP1}
 \end{align}
	\item To avoid the redundant illumination, the $n$th beam position can only be illuminated once during the whole BH period, which can be expressed as
  \begin{align}
	\sum\limits_{t = 1}^M {{x_{n,t}}}  = 1,~\forall n \in {\cal N}.\label{CritriaIP2}
  \end{align}
\end{itemize}

The index set of non-illuminated beam positions at the $t$th time slot, denoted as ${\cal B}$, is initialized to be $\mathcal{N}$. The index set of illuminated beam positions at the $t$th time slot, denoted as ${\cal A}_t$, is determined by randomly choosing $K$ beam positions from ${\cal B}$. At the next time slot, we update ${\cal B}$ by removing the beam positions of ${\cal A}_t$ from ${\cal B}$, which can be expressed as  
\begin{align}\label{IIS_1}
	{{\cal B}} \leftarrow {{\cal B}}\setminus {{\cal A}_{t}}.
\end{align}
By repeating \eqref{IIS_1}, we determine ${\cal A}_{t}$ for $t=1,2,\ldots,M$. Then we can determine $\bd X$ by
	\begin{align}\label{IIS_2}
		{x_{n,t}} = \left\{ \begin{array}{l}
			1,~~~~n \in {{\cal A}_t},t \in {\cal M}\\
			0,~~~~{\rm{otherwise}}
		\end{array} \right..
	\end{align}
Since $\bd X$ satisfies \eqref{CritriaIP1} and \eqref{CritriaIP2}, it is a valid candidate illumination pattern, which can be added to $\left\{{\cal X}_i,i \in {\cal I}\right\}$ by  
	\begin{align}\label{IIS_3}
		{\cal X}_i \leftarrow {\bd X}, {i \in {\cal I}}.
	\end{align}
In this way, we can determine $\left\{{\cal X}_i,i \in {\cal I}\right\}$, which is essentially based on the constrained random search.  In the following, based on the candidate illumination patterns, we will find one achieving the largest sum-rate.
 
%-----------------------------------------------------------------------------------------------------
 Given $\bd{X} = {\cal X}_i,i \in {\cal I}$, the FDBF design subproblem of \eqref{orig_pro1} can be rewritten as
\begin{subequations}\label{P1_orig}
	\begin{align}
		\max_{\left\{ {{{\boldsymbol{P}}_t,t\in\mathcal{M}}} \right\}} &\sum\limits_{n = 1}^{N_{\rm s}} {\sum\limits_{t = 1}^M {{R_{n,t}}\left( {{{\bd{P}}_t},{\bd X}} \right)} } \label{P1_orig_a} \\
		{\rm{s}}{\rm{.t}}{\rm{.}}~~~&\sum\limits_{t = 1}^M {{R_{n,t}}\left( {{{\bd{P}}_t},{\bd X}} \right)}  \ge {{\gamma}_n},\forall n \in {\cal N},\label{P1_orig_b}\\
		&\sum\limits_{n = 1}^{N_{\rm s}} {{{\left\| {{{\boldsymbol{p}}_n^t}} \right\|}_2^2}}  \le {P_{{\rm{tot}}}},\forall t \in {\cal M},\label{P1_orig_c}
	\end{align}
\end{subequations}
%\subsection{FP-based Beamforming Design Method}
which is non-convex due to the logarithmic-fractional form of the objective function \eqref{P1_orig_a} and constraint \eqref{P1_orig_b}. To deal with the non-convexity of \eqref{P1_orig}, the FP method can be applied using the quadratic transform~\cite{Shen_TSP_2018}. To simplify the notation, we define
\begin{equation}
 {{{\widehat{\bd P}}}_k^t} \triangleq {\boldsymbol{p}}_k^t{\left( {{\boldsymbol{p}}_k^t} \right)^{\rm H}}, k\in{\mathcal{N}}.
 \end{equation}
 Then \eqref{P1_orig} can be transformed into
\begin{subequations}\label{P11_1}
	\begin{align}
		\mathop {\max }\limits_{\scriptstyle\left\{ {{{\bd{P}}_t},t \in {\cal M}} \right\},\atop
			\scriptstyle\left\{ {{\mu _{n,t}},t \in {\cal M},n \in {\cal N}} \right\}}  &\sum\limits_{n = 1}^{N_{\rm s}} \sum\limits_{t = 1}^M {{f_{n,t}}\left( {{\boldsymbol{p}}_n^t,{\mu _{n,t}}} \right)}  \label{P11_1a} \\
		{\rm{s}}{\rm{.t}}{\rm{.}}~~~~~~&\sum\limits_{t = 1}^M {{f_{n,t}}\left( {{\boldsymbol{p}}_n^t,{\mu _{n,t}}} \right)}  \ge {{\gamma}_n},\forall n \in {\cal N},\label{P11_1b}\\
		&\eqref{P1_orig_c}.
	\end{align}
\end{subequations}
Note that a new function ${f_{n,t}}\left( {{\boldsymbol{p}}_n^t,{\mu _{n,t}}} \right)$ in \eqref{P11_1} is defined as
\begin{align}
		&{f_{n,t}}\left( {{\boldsymbol{p}}_n^t, {\mu _{n,t}}} \right)  \triangleq \log_2 \Big( 1 + 2{\mathop{\rm Re}\nolimits} \left\{ {\sqrt {{x_{n,t}}} \mu _{n,t}^{\rm H}{\boldsymbol{h}}_n{\boldsymbol{p}}_n^t} \right\} \nonumber\\
			&~~~~~~~~~~~~~~~ -\mu _{n,t}^{\rm H}\Big( {\sum\nolimits_{k \ne n}^{N_{\rm s}} {{{x_{k,t}}{\boldsymbol{h}}_n{{{\widehat{\bd P}}}_k^t}{{\boldsymbol{h}}_n^{\rm H}}}  + {\sigma ^2}}} \Big){\mu _{n,t}} \Big),\label{Func1}		
\end{align}	
where an auxiliary variable $\mu _{n,t}$ is defined as
\begin{equation}\label{AuxiliaryMu}
	\mu  _{n,t}  \triangleq \frac{{\sqrt {{x_{n,t}}} {\boldsymbol{h}}_n{\boldsymbol{p}}_n^t}}{{\sum\nolimits_{k \ne n}^{N_{\rm s}} {{{x_{k,t}}{\boldsymbol{h}}_n{{{\widehat{\bd P}}}_k^t}{{\boldsymbol{h}}_n^{\rm H}}}  + {\sigma ^2}}}}.
\end{equation}
In fact, $\mu _{n,t}$ is determined by setting 
\begin{equation}
	\frac{{\partial {f_{n,t}}\left( {{\bd{p}}_n^t,{\mu _{n,t}}} \right)}}{{\partial {\mu _{n,t}}}} = 0.
\end{equation}

Note that with fixed $\left\{ {{\mu _{n,t}},t \in {\cal M},n \in {\cal N}} \right\}$, \eqref{P11_1} is convex, which is equivalent to \eqref{P1_orig} and can be solved by the interior-point method. By alternately updating $\left\{ {{\mu _{n,t}},t \in {\cal M},n \in {\cal N}} \right\}$ with fixed $\left\{ {{{\bd{P}}_t},t \in {\cal M}} \right\}$ and updating $\left\{ {{{\bd{P}}_t},t \in {\cal M}} \right\}$ with fixed $\left\{ {{\mu _{n,t}},t \in {\cal M},n \in {\cal N}} \right\}$, we run the iteration until triggering \emph{Stop Condition~1}, which can be set as the maximum iteration number being reached. Then we can obtain a solution of $\bd{p}_n^t$ as
\begin{equation}\label{Optimal_p}
	\overline{\bd{p}}_n^t = \arg \mathop { \max }\limits_{{\bd{p}}_n^t } \sum\limits_{n = 1}^{N_{\rm s}} {\sum\limits_{t = 1}^M {{f_{n,t}}\left( {{\bd{p}}_n^t,{\mu _{n,t} }} \right)} }.
\end{equation}	

Therefore, given ${\cal X}_i,i \in {\cal I}$, we can obtain $\overline{\bd{p}}_n^t,~t=1,2,\ldots,M$, based on the procedures from \eqref{P1_orig} to \eqref{Optimal_p}.
Then the designed beamforming matrix given ${\cal X}_i$ is denoted as 
\begin{equation}\label{CandidateP}
	{\overline{\bd{P}}_t^{(i)}} \triangleq [\overline{\bd{p}}_1^t,\overline{\bd{p}}_2^t,\ldots,\overline{\bd{p}}_{N_{\rm s}}^t].
\end{equation}

\begin{algorithm}[!t]
	\caption{FDBF-IPRS Scheme}
	\begin{algorithmic}[1]
		\State \textbf{Input:} $N_{\rm s}$, $N_{\rm BS}$, $K$, $M$ and $I$.
		\State \textbf{Initialization:} ${\cal B} \leftarrow \mathcal{N}$.
		\For{$i = 1,2,...,I$}
		\For{$t = 1,2,...,M$}
		\State Determine ${\cal A}_{t}$ based on ${\cal B}$.
		\State Update ${\cal B}$ via \eqref{IIS_1}.
		\EndFor
		\State Obtain ${\cal X}_i$ via \eqref{IIS_3}.
		\Repeat
		\State Update $\left\{ {{\mu _{n,t}},t \in {\mathcal M},n \in {\mathcal N}} \right\}$ via \eqref{AuxiliaryMu}.
		\State Obtain $\left\{ {{{\bd{P}}_t},t \in {\cal M}} \right\}$ by solving \eqref{P1_orig}.							
		\Until{ \emph{Stop~Condition~1} is triggered}
		\State Obtain ${\overline{\bd{P}}^{(i)}_1},{\overline{\bd{P}}^{(i)}_2},\ldots,{\overline{\bd{P}}^{(i)}_M}$ via \eqref{CandidateP}.
		\EndFor
		\State Obtain $[\widetilde{\bd{P}}_1,\widetilde{\bd{P}}_2,\ldots,\widetilde{\bd{P}}_M, {\widetilde{\bd X}} ]$ via \eqref{Optimal_PX}.
		\State \textbf{Output:} $\widetilde{\bd{P}}_1,\widetilde{\bd{P}}_2,\ldots,\widetilde{\bd{P}}_M$ and ${\widetilde{\bd X}}$.
	\end{algorithmic}
	\label{algorithm2}
\end{algorithm}

Then the optimized $\widetilde{\bd X}$ together with the designed FDBF matrices $\widetilde{\bd{P}}_1,\widetilde{\bd{P}}_2,\ldots,\widetilde{\bd{P}}_M$ can be obtained by
\begin{align}	\label{Optimal_PX}
	&[\widetilde{\bd{P}}_1,\widetilde{\bd{P}}_2,\ldots,\widetilde{\bd{P}}_M, {\widetilde{\bd X}} ] = \nonumber \\
	&~~~~~~~\arg \mathop { \max }\limits_{\left\{{\overline{\bd{P}}^{(i)}_1},{\overline{\bd{P}}^{(i)}_2},\ldots,{\overline{\bd{P}}^{(i)}_M},{\cal X}_i,i \in {\cal I}\right\} } \sum\limits_{n = 1}^{N_{\rm s}} {\sum\limits_{t = 1}^M {{R_{n,t}}\left( {{\overline{\bd{P}}^{(i)}_t},{\cal X}_i} \right)} },
\end{align}
which achieves the largest sum-rate. 

The detailed steps of the proposed FDBF-IPRS scheme are summarized in \textbf{Algorithm~1}. The computational complexity of the FDBF-IPRS scheme is $O\left( I{T_1}{{{\left( M{N_{\rm BS}}{N_{\rm s}} \right)}^{3.5}}{{\log }_2}\left( {1/{\varepsilon_1 } } \right)}  \right)$, where $T_1$ denotes the predefined maximum iteration number of \emph{Stop~Condition~1} and ${\varepsilon }_1 > 0$ is the solution accuracy of the CVX solver used in the FDBF-IPRS scheme~\cite{Grant_2009}.

\subsection{FDBF-IPAO Scheme}\label{subsubsec.FDBF-IPAO}
Due to the binary property of $\bd X$, \eqref{orig_pro1} is a mixed integer non-convex optimization problem, which is difficult to solve. To reduce the difficulty in tackling integer variables, we relax the binary variable $x_{n,t}$ to a continuous variable ${\widehat x}_{n,t}$, with $0 \le {\widehat x}_{n,t} \le 1$, and then transform \eqref{orig_pro1} into a continuous optimization problem. In fact, we will quantize the continuous variables into binary ones after finishing all the iterations.

To solve \eqref{orig_pro1}, we decouple it into two subproblems, namely, the FDBF design subproblem and the illumination pattern design subproblem, and alternately optimize these two subproblems until a stop condition is triggered.

Define $ [\widehat{\bd{X}}]_{n,t} \triangleq {{\widehat x_{n,t}}},~n\in\mathcal{N},~t\in\mathcal{M} $. For the FDBF design subproblem, given $\widehat{\bd{X}}$, \eqref{orig_pro1} can be rewritten as
\begin{subequations}\label{P2_orig}
	\begin{align}
		\max_{\left\{ {{{\boldsymbol{P}}_t,t\in\mathcal{M}}} \right\}} &\sum\limits_{n = 1}^{N_{\rm s}} {\sum\limits_{t = 1}^M {{R_{n,t}}\left( {{{\bd{P}}_t},{\widehat{\bd{X}}}} \right)} } \label{P2_orig_a} \\
		{\rm{s}}{\rm{.t}}{\rm{.}}~~~&\sum\limits_{t = 1}^M {{R_{n,t}}\left( {{{\bd{P}}_t},{\widehat{\bd{X}}}} \right)}  \ge {{\gamma}_n},\forall n \in {\cal N},\label{P2_orig_b}\\
		&\sum\limits_{n = 1}^{N_{\rm s}} {{{\left\| {{{\boldsymbol{p}}_n^t}} \right\|}_2^2}}  \le {P_{{\rm{tot}}}},\forall t \in {\cal M},\label{P2_orig_c}
	\end{align}
\end{subequations}
which is non-convex due to the logarithmic-fractional form of the objective function \eqref{P2_orig_a} and constraint \eqref{P2_orig_b}. To deal with the non-convexity of \eqref{P2_orig}, the FP method can be applied using the quadratic transform~\cite{Shen_TSP_2018}. 
Then \eqref{P2_orig} can be transformed into
\begin{subequations}\label{P2_1}
	\begin{align}
		\mathop {\max }\limits_{\scriptstyle\left\{ {{{\bd{P}}_t},t \in {\cal M}} \right\},\atop
			\scriptstyle\left\{ {{\zeta_{n,t}},t \in {\cal M},n \in {\cal N}} \right\}}  &\sum\limits_{n = 1}^{N_{\rm s}} \sum\limits_{t = 1}^M {{f_{n,t}}\left( {{\boldsymbol{p}}_n^t,{\zeta_{n,t}}} \right)}  \label{P2_1a} \\
		{\rm{s}}{\rm{.t}}{\rm{.}}~~~~~~&\sum\limits_{t = 1}^M {{f_{n,t}}\left( {{\boldsymbol{p}}_n^t,{\zeta_{n,t}}} \right)}  \ge {{\gamma}_n},\forall n \in {\cal N},\label{P2_1b}\\
		&\eqref{P2_orig_c}.
	\end{align}
\end{subequations}
Note that a new function ${f_{n,t}}\left( {{\bd{p}}_n^t,{\zeta_{n,t}}} \right)$ in \eqref{P2_1} is defined as
\begin{align}
	&{f_{n,t}}\left( {{\boldsymbol{p}}_n^t, {\zeta_{n,t}}} \right)  \triangleq \log_2 \Big( 1 + 2{\mathop{\rm Re}\nolimits} \left\{ {\sqrt {\widehat x_{n,t}} \zeta_{n,t}^{\rm H}{\boldsymbol{h}}_n{\boldsymbol{p}}_n^t} \right\} \nonumber\\
	&~~~~~~~~~~~~~~~ -\zeta_{n,t}^{\rm H}\Big( {\sum\nolimits_{k \ne n}^{N_{\rm s}} {{{\widehat x_{k,t}}{\boldsymbol{h}}_n{{{\widehat{\bd P}}}_k^t}{{\boldsymbol{h}}_n^{\rm H}}}  + {\sigma ^2}}} \Big){\zeta _{n,t}} \Big),\label{Func2}		
\end{align}	
where an auxiliary variable $\zeta_{n,t}$ is defined as
\begin{equation}\label{AuxiliaryZeta}
	\zeta_{n,t}  \triangleq \frac{{\sqrt {{\widehat x_{n,t}}} {\bd{h}}_n{\bd{p}}_n^t}}{{\sum\nolimits_{k \ne n}^{N_{\rm s}} {{{\widehat x_{k,t}}{\bd{h}}_n{{{\widehat{\bd P}}}_k^t}{{\bd{h}}_n^{\rm H}}}  + {\sigma ^2}}}}.
\end{equation}
In fact, $\zeta_{n,t}$ is determined by setting 
\begin{equation}
	\frac{{\partial {f_{n,t}}\left( {{\bd{p}}_n^t,{\zeta_{n,t}}} \right)}}{{\partial {\zeta_{n,t}}}} = 0.
\end{equation}

Note that with fixed $\left\{ {{\zeta_{n,t}},t \in {\cal M},n \in {\cal N}} \right\}$, \eqref{P2_1} is convex, which is equivalent to \eqref{P2_orig} and can be solved by the interior-point method. By alternately updating $\left\{ {{\zeta_{n,t}},t \in {\cal M},n \in {\cal N}} \right\}$ with fixed $\left\{ {{{\bd{P}}_t},t \in {\cal M}} \right\}$ and updating $\left\{ {{{\bd{P}}_t},t \in {\cal M}} \right\}$ with fixed $\left\{ {{\zeta_{n,t}},t \in {\cal M},n \in {\cal N}} \right\}$, we run the iteration until triggering \emph{Stop Condition~2}, which can be set as the maximum iteration number being reached. Then we obtain a solution of $\bd{p}_n^t$ as
\begin{equation}\label{Optimal_p2}
	\overrightarrow{\bd{p}}_n^t = \arg \mathop { \max }\limits_{{\bd{p}}_n^t } \sum\limits_{n = 1}^{N_{\rm s}} {\sum\limits_{t = 1}^M {{f_{n,t}}\left( {{\bd{p}}_n^t,{\zeta_{n,t} }} \right)} }.
\end{equation}	

Therefore, given $\widehat{\bd{X}}$, we can obtain $\overrightarrow{\bd{p}}_n^t,~t=1,2,\ldots,M$, based on the procedures from \eqref{P2_orig} to \eqref{Optimal_p2}.
Then the designed FDBF matrix given $\widehat{\bd X}$ is denoted as 
\begin{equation}\label{RightArrowP}
	\overrightarrow{\bd{P}}_t \triangleq [\overrightarrow{\bd{p}}_1^t,\overrightarrow{\bd{p}}_2^t,\ldots,\overrightarrow{\bd{p}}_{N_{\rm s}}^t].
\end{equation} 

For the illumination pattern design subproblem, given $\overrightarrow{\bd{P}}_1,\overrightarrow{\bd{P}}_2,...,\overrightarrow{\bd{P}}_M$, \eqref{orig_pro1} can be rewritten as
\begin{subequations}\label{FDBF-IP-AO.Problem}
	\begin{align}
		\max_{\widehat{\bd{X}}}& \sum\limits_{n = 1}^{N_{\rm s}} {\sum\limits_{t = 1}^M {{R_{n,t}}\left( {\overrightarrow{\bd{P}}_t,{\widehat{\bd{X}}}} \right)} } \label{FDBF-IP-AO.Problem_a}\\
		{\rm{s}}{\rm{.t}}{\rm{.}}&\sum\limits_{t = 1}^M {{R_{n,t}}\left( {\overrightarrow{\bd{P}}_t,{\widehat{\bd{X}}}} \right)}  \ge {{\gamma}_n},\forall n \in {\cal N},\label{FDBF-IP-AO.Problem_b}\\
		&\sum\limits_{n = 1}^{N_{\rm s}} {{x_{n,t}}}  \le K,\forall t \in {\cal M},\label{FDBF-IP-AO.Problem_c}\\
		&{x_{n,t}} \in \left\{ {0,1} \right\},\forall n\in {\cal N},\forall t \in {\cal M}.\label{FDBF-IP-AO.Problem_d}
	\end{align}
\end{subequations}

Note that \eqref{FDBF-IP-AO.Problem} is non-convex because of the logarithmic-fractional form of the objective function \eqref{FDBF-IP-AO.Problem_a} and the constraint \eqref{FDBF-IP-AO.Problem_b}. To deal with the non-convexity of \eqref{FDBF-IP-AO.Problem}, we define an auxiliary variable as
\begin{align}\label{AuxiliaryXi}
	\xi _{n,t}  \triangleq \frac{{\sqrt {{{\widehat x}_{n,t}}} {{\bd{h}}_n^{\rm H}{\overrightarrow{\bd{p}}_n^t}}}}{{\sum\nolimits_{k \ne n} {{{\widehat x}_{k,t}}{\bd{h}}_n{{{\widecheck{\bd P}}}_k^t}{{\bd{h}}_n^{\rm H}}}  + {\sigma ^2}}},
\end{align}
where 
\begin{equation}
	{{\widecheck{\bd P}}_k^t} \triangleq {\overrightarrow{\bd{p}}_k^t}{\left( {\overrightarrow{\bd{p}}_k^t} \right)^{\rm H}}, k\in{\mathcal{N}}.
\end{equation}
Then we define two vectors ${\bd{v}}_n^t \in \mathbb{C}^{M{N_{\rm s}}}$ and ${\bd{d}}_n^t \in \mathbb{C}^{M{N_{\rm s}}}$ respectively as
\begin{align}
	&{\bd{v}}_n^t \triangleq  2{\mathop{\rm Re}\nolimits} \left\{ {\xi _{n,t}^{\rm H}{\bd{h}}_n{\overrightarrow{\bd{p}}_n^t}} \right\}{{\bd{e}}_M^t} \otimes {{\bd{e}}_{N_{\rm s}}^n}, \\
	& {\bd{d}}_n^t \triangleq  \bd{e}_M^t \otimes {\widehat{\bd{d} }_n^t},
\end{align}
where $\bd{e}_M^t$ denotes an $M$-dimensional column vector with its entries all being zero except its $t$th entry being one, $\bd{e}_{N_{\rm s}}^n$ denotes an $N_{\rm s}$-dimensional column vector with its entries all being zero except its $n$th entry being one, and the $i$th entry of ${\widehat {\bd{d}}_n^t}$ is
\begin{align}
	{\left[ {\widehat {\bd{d}}_n^t} \right]_i} \triangleq \left\{ \begin{array}{l}
		{\bd{h}}_n{{{\widecheck{\bd P}}}_k^t}{{\bd{h}}_n^{\rm H}},~~~~~~i\in \mathcal{N}, i \ne k,\\
		0,~~~~~~~~~~~~~~~~i = k.
	\end{array} \right.
\end{align}
To give compact notation, we stack the relaxed continuous variables together as
\begin{align}
	{\widehat{\boldsymbol{x}}_t} \triangleq & {\left[ {{{\widehat x}_{1,t}},{{\widehat x}_{2,t}},...,{{\widehat x}_{{N_{\rm s}},t}}} \right]^{\rm T}}, \\
	{\boldsymbol{x}} \triangleq & {\left[ {{\widehat{\boldsymbol{x}}}_1^{\rm T},{\widehat{\boldsymbol{x}}}_2^{\rm T},...,{\widehat{\boldsymbol{x}}}_M^{\rm T}} \right]^{\rm T}}.\label{RelaxedX}
\end{align}

Based on the quadratic transform, the achievable rate in \eqref{FDBF-IP-AO.Problem_a} can be expressed as
\begin{align}
	R_{n,t}({\overrightarrow{\bd{P}}_t,{\widehat{\bd{X}}}}) & =  \log_2 \left( {1 + {\sqrt{\bd{x}}^{\rm T}}{\bd{v}}_n^t - \xi _{n,t}^{\rm H}\left( {{{\bd{x}}^{\rm T}}{\bd{d}}_n^t + {\sigma ^2}} \right){\xi _{n,t}}} \right) \nonumber \\
	&\triangleq {g_{n,t}}\left( { \xi_{n,t},  \bd{x} } \right), 
\end{align}
where $\sqrt {\bd x}$ is a vector with the same dimension as $\bd x$ and each entry of $\sqrt {\bd x}$ is the square root of that of ${\bd{x}}$, and ${g_{n,t}}\left( { \xi_{n,t},  \bd{x} } \right)$ is a newly defined function. In fact, $\xi _{n,t} $ in \eqref{AuxiliaryXi} can be obtained by setting
\begin{equation}
	\frac{{\partial {g_{n,t}}\left( { \xi_{n,t},  \bd{x} } \right)}}{{\partial {\xi _{n,t}}}} = 0.
\end{equation}

Then \eqref{FDBF-IP-AO.Problem} can be transformed into
\begin{subequations}\label{FDBF-IP-AO.IPProblem}
	\begin{align}
		\mathop {\max }\limits_{ {\boldsymbol{x}},\left\{ {{\xi _{n,t}}}, n\in \mathcal{N}, t\in \mathcal{M} \right\}} &\sum\limits_{n = 1}^{N_{\rm s}} {\sum\limits_{t = 1}^M {g_{n,t}}\left( { \xi_{n,t},  \bd{x} } \right) }\label{FDBF-IP-AO.IPProblem_a}\\
		{\rm{s}}.{\rm{t}}.~~~~~~~&\sum\limits_{t = 1}^M {g_{n,t}}\left( { \xi_{n,t},  \bd{x} } \right)  \ge {\gamma _n},\label{FDBF-IP-AO.IPProblem_b}\\
		&\left( {{{\bd I}_M} \otimes {\bd 1}_{N_{\rm s}}^{\rm T}} \right){\bd{x}} \preceq K \cdot {\bd 1}_M,\label{FDBF-IP-AO.IPProblem_c}\\
		&{\bd A}{\bd{x}} \succeq {\bd 1}_{N_{\rm s}},\label{FDBF-IP-AO.IPProblem_d}
	\end{align}
\end{subequations}
where $\bd A \triangleq [\bd I_{N_{\rm s}},\bd I_{N_{\rm s}},...,\bd I_{N_{\rm s}}] \in {\mathbb R}^{{N_{\rm s}}\times M{N_{\rm s}}}$,  ${\bd 1}_{N_{\rm s}}$ and ${\bd 1}_M$ denote the ${N_{\rm s}}$-dimensional and $M$-dimensional column vector with its entries all being one, respectively.

Note that with fixed $\left\{ {{\xi _{n,t}}}, n\in \mathcal{N}, t\in \mathcal{M} \right\}$, \eqref{FDBF-IP-AO.IPProblem} is convex, which can be solved by the existing convex optimization tools. By alternately updating $\left\{ {{\xi _{n,t}}}, n\in \mathcal{N}, t\in \mathcal{M} \right\}$ with a fixed $\bd{x}$ and updating $\bd{x}$ with fixed $\left\{ {{\xi _{n,t}}}, n\in \mathcal{N}, t\in \mathcal{M} \right\}$, we run the iterations until triggering \emph{Stop Condition~3}, which can be set as the maximum iteration number being reached. The computational complexity of solving \eqref{FDBF-IP-AO.IPProblem} is $O\Big( {T_3}{{{\left( M{N_{\rm s}} \right)}^{3.5}}{{\log }_2}\left( {1/{\varepsilon_2} } \right)} \Big)$, where $T_3$ denotes the predefined maximum iteration number of \emph{Stop~Condition~3} and ${\varepsilon }_2 > 0$ is the solution accuracy of the CVX solver~\cite{Grant_2009}.

\begin{algorithm}[!t]
	\caption{FDBF-IPAO Scheme}
	\begin{algorithmic}[1]
		\State \textbf{Input:} $N_{\rm s}$, $N_{\rm BS}$, $K$, $M$.
		\State \textbf{Initialization:} Initialize $\boldsymbol{x}$ via \eqref{Initialize_x}.
		\Repeat
		\Repeat		
		\State Update  $\left\{ {{\zeta_{n,t}},t \in {\mathcal M},n \in {\mathcal N}} \right\}$  via \eqref{AuxiliaryZeta}.		
        \State Obtain $\left\{ {{{\bd{P}}_t},t \in {\cal M}} \right\}$ by solving \eqref{P2_orig}.
        \Until{ \emph{Stop Condition~2} is triggered}		
		\State Obtain $\overrightarrow{\bd{P}}_1,\overrightarrow{\bd{P}}_2,\ldots,\overrightarrow{\bd{P}}_M $ via \eqref{RightArrowP}.
		\Repeat
		\State Update  $\left\{ {{\xi _{n,t}},t \in {\mathcal M},n \in {\mathcal N}} \right\}$  via \eqref{AuxiliaryXi}.		
		\State Obtain $\bd x$ by solving \eqref{FDBF-IP-AO.IPProblem}.
		\Until{ \emph{Stop Condition~3} is triggered}
		\Until{ \emph{Stop Condition~4} is triggered}
		\State  Obtain $\widetilde{\bd X}$ via \eqref{Quantization}.
		\State Substitute $\widehat{\bd X}$ in \eqref{P2_orig} by $\widetilde{\bd X}$ and then solve \eqref{P2_orig}, obtaining $\overrightarrow{\bd{P}}_1,\overrightarrow{\bd{P}}_2,\ldots,\overrightarrow{\bd{P}}_M $.
%		\State Obtain $\overrightarrow{\bd{F}}_1,\overrightarrow{\bd{F}}_2,\ldots,\overrightarrow{\bd{F}}_M$ and  $\overrightarrow{\bd{Q}}_1,\overrightarrow{\bd{Q}}_2,\ldots,\overrightarrow{\bd{Q}}_M$ by employing \textbf{Algorithm~1}.
		\State \textbf{Output:} $\overrightarrow{\bd{P}}_1,\overrightarrow{\bd{P}}_2,\ldots,\overrightarrow{\bd{P}}_M $, $\widetilde{\bd X}$.
	\end{algorithmic}
	\label{algorithm3}
\end{algorithm} 

Based on the solutions of the FDBF design subproblem and illumination pattern design subproblem, now we propose an FDBF-IPAO scheme for the fully-digital BIP design problem by alternately optimizing these two subproblems until a stop condition is triggered. 

Specifically, we initialize a feasible $\bd x$ by setting all its entries being $K/{N_{\rm s}}$ to satisfy~\eqref{FDBF-IP-AO.IPProblem_c} and \eqref{FDBF-IP-AO.IPProblem_d}, i.e.,
\begin{equation}\label{Initialize_x}
	[\bd{x}]_i = K/{N_{\rm s}},~i=1,2,\ldots,M{N_{\rm s}}.
\end{equation}
%{\color{red}where $\bd x$ with its entries all being equal may not maximize the strength of desired signals, but it poses a basic trend for the beamforming design.} 
Given $\bd x$, we update $\left\{ {{\zeta_{n,t}},t \in {\mathcal M},n \in {\mathcal N}} \right\}$ via \eqref{AuxiliaryZeta} and then obtain $\left\{ {{{\bd{P}}_t},t \in {\cal M}} \right\}$ by solving \eqref{P2_orig}, which is indicated by step 4 to step 7 and is iteratively executed until triggering \emph{Stop Condition~2}. Then we obtain $\overrightarrow{\bd{P}}_1,\overrightarrow{\bd{P}}_2,\ldots,\overrightarrow{\bd{P}}_M$  via  \eqref{RightArrowP}. Based on the obtained $\overrightarrow{\bd{P}}_1,\overrightarrow{\bd{P}}_2,\ldots,\overrightarrow{\bd{P}}_M$, we updated  $\left\{ {{\xi _{n,t}},t \in {\mathcal M},n \in {\mathcal N}} \right\}$ via \eqref{AuxiliaryXi} and then obtain $\bd x$ by solving \eqref{FDBF-IP-AO.IPProblem}, which is indicated by step 9 to step 12 and is iteratively performed until triggering \emph{Stop Condition~3}. We alternately run the above steps until triggering \emph{Stop Condition~4}, which can be simply set that a predefined maximum number of iteration is reached. In fact, \emph{Stop Condition~4} can be more effectively set that the objective of \eqref{P2_orig} can no longer increase.

To satisfy \eqref{FDBF-IP-AO.Problem_d}, we need to quantize the continuous variables into binary ones by
\begin{equation}\label{Quantization}
	\big[ {\widetilde{\bd X}}\big] _{n,t} = \lfloor {\bd x}_{n+{N_{\rm s}}\left(t-1\right)} +0.5 \rfloor.
\end{equation}
If $\sum\nolimits_{i = 1}^{N_{\rm s}} \big[{\widetilde{\bd X}}\big]_{i,t}  > K,t\in{\cal M}$, we define $\big[{\widetilde{\bd X}}\big]_{n,t} = {\bd x}_{n+{N_{\rm s}}\left(t-1\right)}$ and set the largest $K$ entries in the $t$th column of $\widetilde{\bd X}$ as $1$. Then, we achieve the designed illumination pattern $\widetilde{\bd X}$. Then we substitute $\widehat{\bd X}$ in \eqref{P2_orig} by $\widetilde{\bd X}$ and then solve \eqref{P2_orig}, obtaining $\overrightarrow{\bd{P}}_1,\overrightarrow{\bd{P}}_2,\ldots,\overrightarrow{\bd{P}}_M $ as the finally designed fully-digital beamformers.

%\footnote{If $\sum\nolimits_{i = 1}^{N_{\rm s}} \big[{\widetilde{\bd X}}\big]_{i,t}  \ge K,\forall t \in {\cal M}$, we define $\big[{\widetilde{\bd X}}\big]_{n,t} = {\bd x}_{n+{N_{\rm s}}\left(t-1\right)}$ and then set the largest $K$ entries in the $t$th column of $\widetilde{\bd X}$ as one.}
%If $\sum\nolimits_{i = 1}^{N_{\rm s}} \big[{\widetilde{\bd X}}\big]_{i,t}  \ge K$ for $t=1,2,...,M$, we define $\big[{\widetilde{\bd X}}\big]_{n,t} = {\bd x}_{n+{N_{\rm s}}\left(t-1\right)}$ and set the largest $K$ entries in each column of $\widetilde{\bd X}$ as one. Then, we achieve the designed illumination pattern $\widetilde{\bd X}$.
%which is the designed illumination pattern.

The detailed steps of the proposed FDBF-IPAO scheme are summarized in \textbf{Algorithm~2}. The computational complexity of FDBF-IPAO scheme is $O\big( \left( \left( {T_4+1} \right) {T_2}{N_{\rm BS}^{3.5}} + {T_4}{T_3}\right)\left( {MN_{\rm s}}\right) ^{3.5}  {\log }_2\big( {1/{\varepsilon_2} } \big) \big)$, where $T_2$ and $T_4$ denote the maximum iteration numbers of \emph{Stop Condition 2} and \emph{Stop Condition 4}, respectively.

\subsection{HBF-AM algorithm}\label{subsection.HBF-AM}
Based on the FDBF matrices $\widetilde{\bd{P}}_1$, $\widetilde{\bd{P}}_2$, \ldots, $\widetilde{\bd{P}}_M$ designed by the FDBF-IPRS scheme or $\overrightarrow{\bd{P}}_1$, $\overrightarrow{\bd{P}}_2$, \ldots, $\overrightarrow{\bd{P}}_M $ designed by the FDBF-IPAO scheme, now we determine the hybrid beamformers, which are essentially the multiplication of the analog beamformers and digital beamformers. Without loss of generality, we use the designed FDBF matrices $\widetilde{\bd{P}}_1$, $\widetilde{\bd{P}}_2$, \ldots, $\widetilde{\bd{P}}_M$ to present the HBF-AM algorithm.

By minimizing the Euclidean distance between the designed fully-digital beamformers and the hybrid beamformers, the HBF design problem for $t =1,2,...,M$ can be expressed as  
\begin{subequations}\label{AlterMinProblem}
	\begin{align}
		\min_{{\bd F}_t,{\bd Q}_t}~&{\big\| {\widetilde{\bd{P}}_t}-{{\bd F}_t}{{\bd Q_t}}\big\|}_{\rm F}  \label{AlterMinProblem_a} \\
		{\rm {s.t.}}~~&{\left\|{{\bd F}_t}{{\bd Q_t}}\right\|}_{\rm F}^2\le {P_{\rm{tot}}},\label{AlterMinProblem_b}\\
		&\left| {[{{\bd{F}}_t}]_{i,n}} \right| = 1,\forall i \in {{\cal N}_{{\rm{BS}}}},n \in {{\cal N}},\label{AlterMinProblem_c}
	\end{align}
\end{subequations}
where \eqref{AlterMinProblem_b} is the total transmission power constraint and \eqref{AlterMinProblem_c} is the unit-modulus constraint due to the phase shifters in analog beamformers. In fact, we can temporarily neglect \eqref{AlterMinProblem_b} to solve the following problem as 
\begin{subequations}\label{AlterMinProblem2}
	\begin{align}
		\min_{{\bd F}_t,{\bd Q}_t}~&{\big\| {\widetilde{\bd{P}}_t}-{{\bd F}_t}{{\bd Q_t}}\big\|}_{\rm F}  \label{AlterMinProblem2_a} \\
		{\rm {s.t.}}~~&\left| {[{{\bd{F}}_t}]_{i,n}} \right| = 1,\forall i \in {{\cal N}_{{\rm{BS}}}},n \in {{\cal N}}.\label{AlterMinProblem2_c}
	\end{align}
\end{subequations}
Note that \eqref{AlterMinProblem_b} can be satisfied via normalizing ${\bd Q}_t$ obtained from \eqref{AlterMinProblem2}. For~\eqref{AlterMinProblem2}, the analog beamformers $\bd{F}_t$ and digital beamformers $\bd{Q}_t$ can be designed by alternately performing the following two steps.

\begin{itemize}
	
	\item{\textit{Step~1:}} Given a feasible solution of ${\bd F}_t$ denoted as ${\widehat{\bd F}}_t$, the digital beamformer design problem can be formulated as
	\begin{align}\label{OptimalQtProblem}
		\min_{{\bd Q}_t}~&{\big\| {\widetilde{\bd{P}}_t}-{{\bd F}_t}{{\bd Q_t}}\big\|}_{\rm F}  
	\end{align}
	which can be solved by the least squared method with the solution as
	\begin{align}\label{OptimalQt}
		{{\widehat{\bd Q}}_t} = \left( {\bd F}_t^{\rm H}{\bd F}_t\right) ^{-1}{{\bd F}_t^{\rm H}}{\widetilde{\bd{P}}_t}.
	\end{align}
	\item{\textit{Step~2:}}  Given $\widehat {\bd Q}_t$, the analog beamformer design problem can be formulated as 
	\begin{subequations}\label{OptimalFtProblem}
		\begin{align}
			\min_{{\bd F}_t}~&{\big\| {\widetilde{\bd{P}}_t}-{{\bd F}_t}{\widehat {\bd Q}_t}\big\|}_{\rm F}  \label{OptimalFtProblem_a} \\
			{\rm {s.t.}}~~&\left| {[{{\bd{F}}_t}]_{i,n}} \right| = 1,\forall i \in {{\cal N}_{{\rm{BS}}}},n \in {{\cal N}},\label{OptimalFtProblem_b}
		\end{align}
	\end{subequations}
	which can be solved by the Riemannian manifold optimization method using the existing toolbox. 
\end{itemize}

We iteratively perform the above two steps until \textit{Stop~Condition~5} is triggered, which can be set as the maximum iteration number being reached. For the obtained ${\widehat{\bd Q}}_t$ and ${\widehat{\bd F}}_t$ after finishing all the iterations, we normalize ${\widehat{\bd Q}}_t$ to satisfy \eqref{AlterMinProblem_b} via
\begin{align}\label{OptimalFt}
	{\widehat{\bd Q}}_t \leftarrow \frac{\sqrt {P_{\rm {tot}}}}{\big\| {{\widehat{\bd F}}_t}{{\widehat{\bd Q}}_t}\big\|_{\rm F}}{{\widehat{\bd Q}}_t}.
\end{align}

\begin{algorithm}[!t]
	\caption{HBF-AM algorithm}
	\begin{algorithmic}[1]
		\State \textbf{Input:} $\widetilde{\bd{P}}_1,\widetilde{\bd{P}}_2,\ldots,\widetilde{\bd{P}}_M$
		\For{$t =1,2,\ldots,M$}
		\Repeat
		\State Randomly generate ${\bd F}_t$ based on \eqref{AlterMinProblem_c}.
		\State Compute ${\widehat {\bd Q}_t}$ via \eqref{OptimalQt}.
		\State Obtain ${\widehat {\bd F}_t}$ by solving \eqref{OptimalFtProblem}.
		\Until{ \emph{Stop~Condition~5} is triggered}
		\State Normalized ${\widehat {\bd Q}_t}$ via \eqref{OptimalFt}.
		\EndFor
		%		\State \textbf{Output:} $\widehat{\bd{F}}_1,\widehat{\bd{F}}_2,\ldots,\widehat{\bd{F}}_M$, $\widehat{\bd{Q}}_1,\widehat{\bd{Q}}_2,\ldots,\widehat{\bd{Q}}_M$.
		\State \textbf{Output:} ${\widehat{\bd F}}_1,{\widehat{\bd F}}_2,\ldots,{\widehat{\bd F}}_M$ and ${\widehat{\bd Q}}_1,{\widehat{\bd Q}}_2,\ldots,{\widehat{\bd Q}}_M$.
	\end{algorithmic}
	\label{algorithm1}
\end{algorithm} 

We repeat the same procedure for $t=1,2,\ldots,M$ and obtain the analog beamformers $\widehat{\bd{F}}_1,\widehat{\bd{F}}_2,\ldots,\widehat{\bd{F}}_M$ and  digital beamformers $\widehat{\bd{Q}}_1,\widehat{\bd{Q}}_2,\ldots,\widehat{\bd{Q}}_M$. 

The detailed steps for the HBF-AM algorithm are summarized in \textbf{Algorithm~3}. Note that the computational complexity of the HBF-AM algorithm is $O\big( M{T_5}{\big( N_{\rm s}^3+{N_{\rm BS}}{N_{\rm s}} \big)} \big)$, where $T_5$ denotes the predefined maximum iteration number of \emph{Stop~Condition~5}.

To solve the joint BIP problem in \eqref{orig_pro}, if we use the FDBF-IPRS scheme followed by the HBF-AM algorithm,  
the computational complexity is 
\begin{equation}
	O\left( I{T_1}{{{\left( M{N_{\rm BS}}{N_{\rm s}} \right)}^{3.5}}{{\log }_2}\left( {1/{\varepsilon_1} } \right)}  +M{T_5}{\left( N_{\rm s}^3+{N_{\rm BS}}{N_{\rm s}} \right)} \right).
\end{equation}
If we use the FDBF-IPAO scheme followed by the HBF-AM algorithm, the computational complexity is 
\begin{align}\label{ComplexityAlgorithm5}
	&O\Big( \left( \left( {T_4+1} \right) {T_2}{N_{\rm BS}}^{3.5} + {T_4}{T_3}\right)\left( {MN_{\rm s}}\right) ^{3.5}  {\log }_2\big( {1/{\varepsilon_2} } \big)\Big.\nonumber \\ 	
	&~~~~~~~~~~~~~~~~~~~~~~~~~~~~~~~~~\Big.+ M{T_5}{\big( N_{\rm s}^3+{N_{\rm BS}}{N_{\rm s}} \big)} \Big).
\end{align}
%Since the computational complexity to determine the fully-digital beamformers given the illumination pattern based on the FP method are similar for the FDBF-IPRS and FDBF-IPAO schemes, $T_1$ and $T_2$ are similar. To find a near-optimal solution from the randomly generated illumination pattern, $I$ needs to be set large, where $I$ is typically much larger than $T_4$. Therefore, the computational complexity of FDBF-IPRS is much higher than that of FDBF-IPAO. On the other hand, as $I$ grows to be infinity, the optimal solution will be obtained by FDBF-IPRS.
Since the computational complexity to determine the fully-digital beamformers given the illumination pattern based on the FP method is similar for the FDBF-IPRS and FDBF-IPAO schemes, $T_1$ and $T_2$ are similar. To find a near-optimal solution from the randomly generated illumination pattern, $I$ needs to be set large, where $I$ is typically much larger than $T_4$. To be specific, we set 
		\begin{equation}
			I=\frac{1}{{M!}} \underbrace {C_{{N_{\rm s}}}^K C_{{N_{\rm s}} - K}^K \cdots C_{{N_{\rm s}} - \left( {M - 1} \right)K}^K}_M ,
		\end{equation}
		where $C_{{N_{\rm s}}}^K$ denotes the binomial coefficient of choosing $K$ beam positions from the total $N_{\rm s}$ beam positions. It is seen that as $N_{\rm s}$ becomes large, $I$ gets extremely large. Therefore, the computational complexity of FDBF-IPRS is much higher than that of FDBF-IPAO. On the other hand, as $I$ grows to be infinity, the optimal solution will be obtained by FDBF-IPRS.

%-----------------------------------------------------------------------------------------------------
\section{Simulation Results}
\begin{table}[!t]  
	\begin{center}
		\renewcommand{\arraystretch}{1.2}   
		\caption{Simulation parameters} \label{table} 
		\fontsize{9}{12}\selectfont  
		\begin{tabular}{cc}% 其中，tabular是表格内容的环境；c表示centering，即文本格式居中；c的个数代表列的个数
			\hline\hline\noalign{\smallskip}     
			\textbf{Parameter} & \textbf{Value} \\ %换行
			\midrule %[2pt]  
			Numbers of satellite antennas $N_{\rm BS}$ & 64  \\
			Numbers of total beam positions $N_{\rm s}$ & 6 \\
			Numbers of time slots $M$ & 3\\
			Rician factor ${\chi_l^{\left(n\right)}}$ & 10 dB \\
			Bandwidth $B$ & 250 MHz\\
			Carrier frequency $f_c$& 20 GHz\\
			Boltzmann’s constant $\kappa$ & $1.38 \times 10^{-23} {\rm J}\cdot {\rm K}^{-1}$\\
			Receiving noise temperature $T_{\rm R}$ & 293 K\\
			\midrule %[2pt]     
		\end{tabular}
	\end{center}   
\end{table}
We assume that the LEO satellite is equipped with $N_{\rm BS} = 64$ antennas, $N_{\rm RF} =6$ RF chains and $N_{\rm s}=6$ beam positions, where the maximum number of illuminated beam positions is $K=2$ and the number of BH time slots is $M=3$. The channels between the LEO satellite and each beam position are supposed to include $L_n=2$ channel paths, where the channel gain $g_l^{\left(n\right)}$  obeys Rician fading distribution with Rician factor ${\chi  _l^{\left(n\right)}} = 10$~dB. The simulation parameters are mainly summarized in Table~\ref{table}. Note that in this paper the HBF design is performed by the HBF-AM algorithm, based on the FDBF designed by either the FDBF-IPRS or FDBF-IPAO scheme. For simplicity, we set $N_{\rm s}=KM$. In fact, if $N_{\rm s}$ cannot be evenly divided by $M$ or $K$, we suppose $N_{\rm s}=KM + D$, where $N_{\rm s}$ is divided by $K$ with a remainder $D$. Then we need $M+1$ time slots, where $D$ beam positions are illuminated in one time slot  and $K$ beam positions are illuminated in each of the $M$ time slots.

\begin{figure}[!t]
	\includegraphics[width=0.5\textwidth]{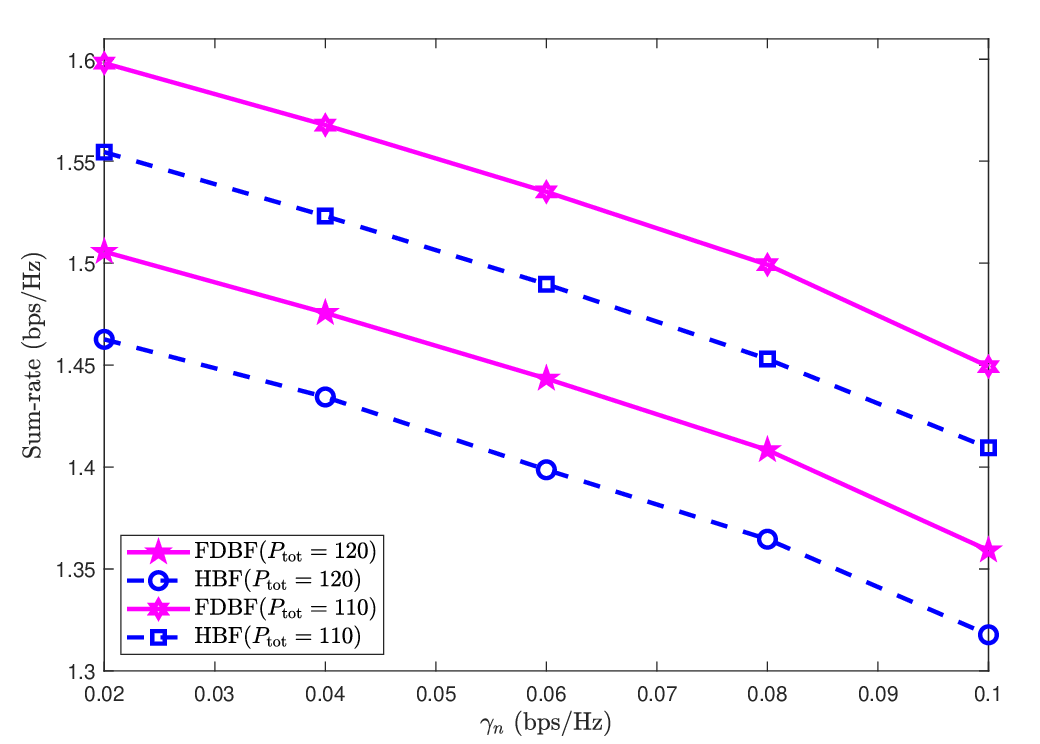}
	\caption{Comparisons of sum-rate of total $N_{\rm s}$ beam positions under different $\gamma_n$ for FDBF design and HBF design with different $P_{\rm tot}$.~~~~~~~~~~~~~~~~~~~~~}
	\label{RthredDiffSchemes.Fig}
\end{figure}

\begin{figure}[!t]
	\includegraphics[width=0.5\textwidth]{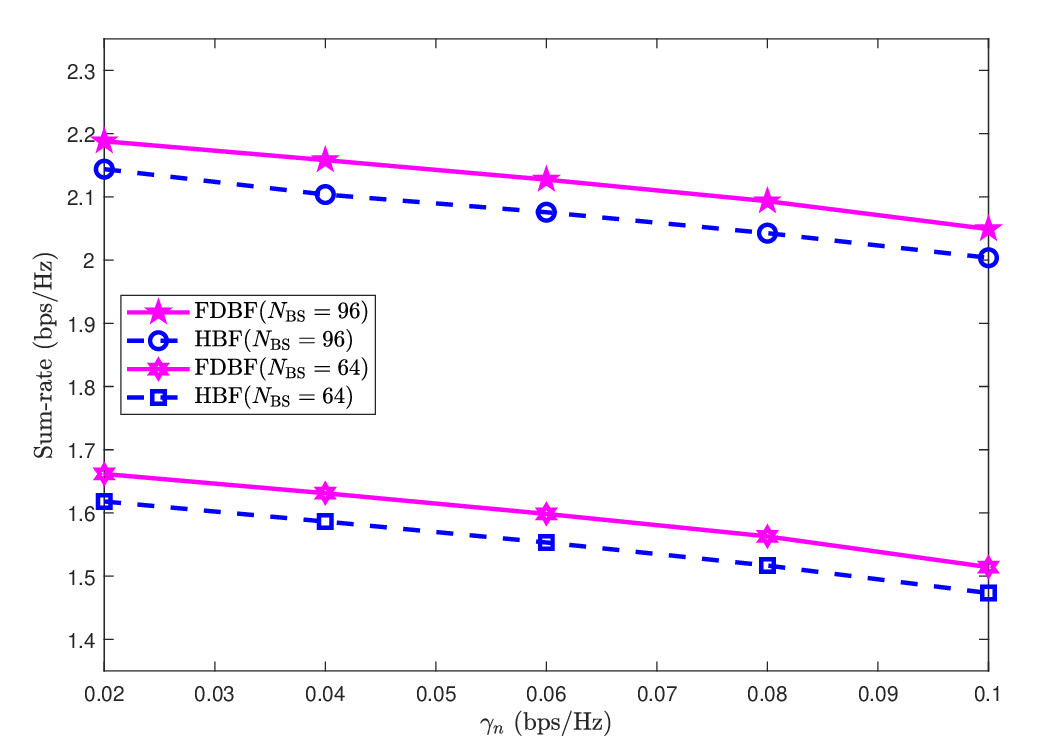}
	\caption{Comparisons of sum-rate of total $N_{\rm s}$ beam positions under different $N_{\rm BS}$ for FDBF design and HBF design with different $\gamma_n$.~~~~~~~~~~~~~~~~~~~~~~}
	\label{NbsNumberRthred.Fig}
\end{figure}

As shown in Fig.~\ref{RthredDiffSchemes.Fig}, we compare the sum-rate of total $N_{\rm s}$ beam positions  under  different $\gamma_n$ and  $P_{\rm tot}$ with fixed $N_{\rm BS} = 64$ and $N_{\rm s} = 6$, where the FDBF-IPRS scheme is employed for both the FDBF and HBF design. For simplicity, $\gamma_n$ is set the same for different $n$.  It is observed that as $\gamma_n$ increases, the sum-rate of total $N_{\rm s}$ beam positions falls, which implies that a high requirement of the sum-rate from the individual beam position results in a poor sum-rate of the total beam positions.  When $P_{\rm tot}$ grows from $110$ to $120$~W, the sum-rate can be increased by around 10\% for either FDBF and HBF given the same $\gamma_n$.

As shown in Fig.~\ref{NbsNumberRthred.Fig}, we compare the sum-rate of total $N_{\rm s}$ beam positions under different $\gamma_n$ and  $N_{\rm BS}$ with fixed $N_{\rm s} = 6$ and $P_{\rm tot} = 120$~W, where the FDBF-IPRS scheme is employed
for both the FDBF and HBF design. For simplicity, we set $\gamma_n$ the same for different $n$. It can be observed that the sum-rate of total $N_{\rm s}$ beam positions falls as $\gamma_n$ increases. For both FDBF and HBF with the same $\gamma_n$, a larger antenna array provides more beamforming design degrees of freedom. When increasing from $N_{\rm BS} = 64$ to $N_{\rm BS} = 96$, the sum-rate can be improved by around 33\%. 

\begin{figure}[!t]
	\includegraphics[width=0.5\textwidth]{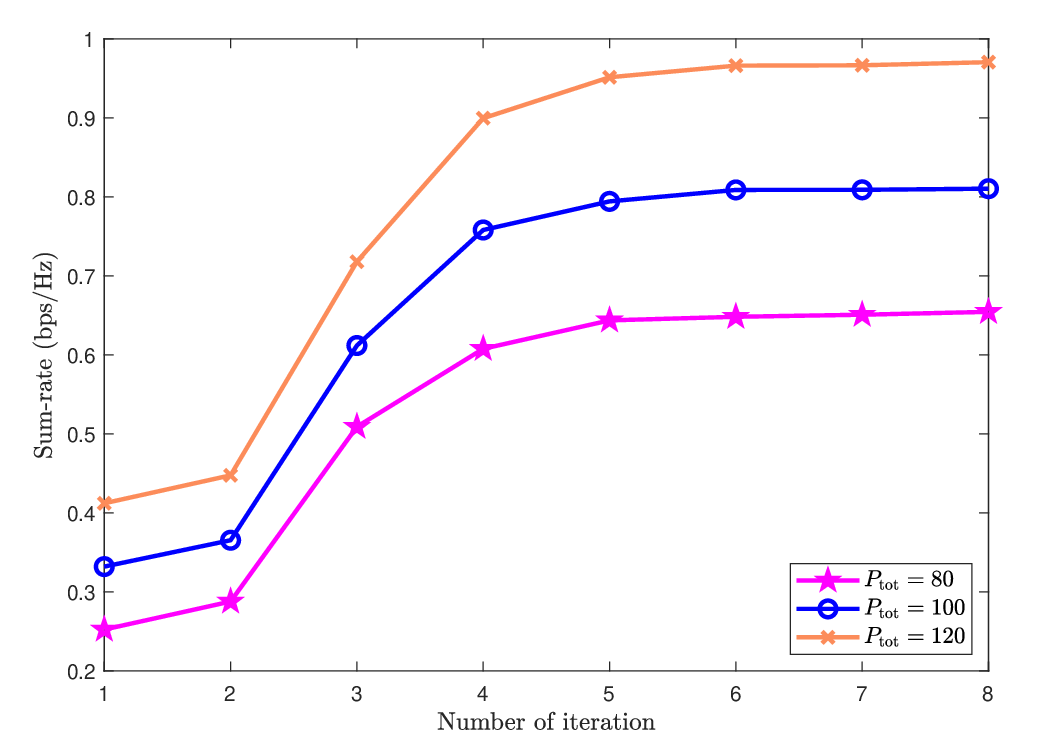}
	\caption{Convergence of the sum-rate of FDBF-IPAO scheme with different $P_{\rm tot}$.~~~~~~~~~~~~~~~~~~~~~~~~~~~~~~~~~~~~~~~~~~~~~~~~~~~~~~~~~~~~~~~~~~~~~~~~~~~~~~~~~~~}
	\label{Iteration.Fig}
\end{figure}

\begin{figure}[!t]
	\includegraphics[width=0.5\textwidth]{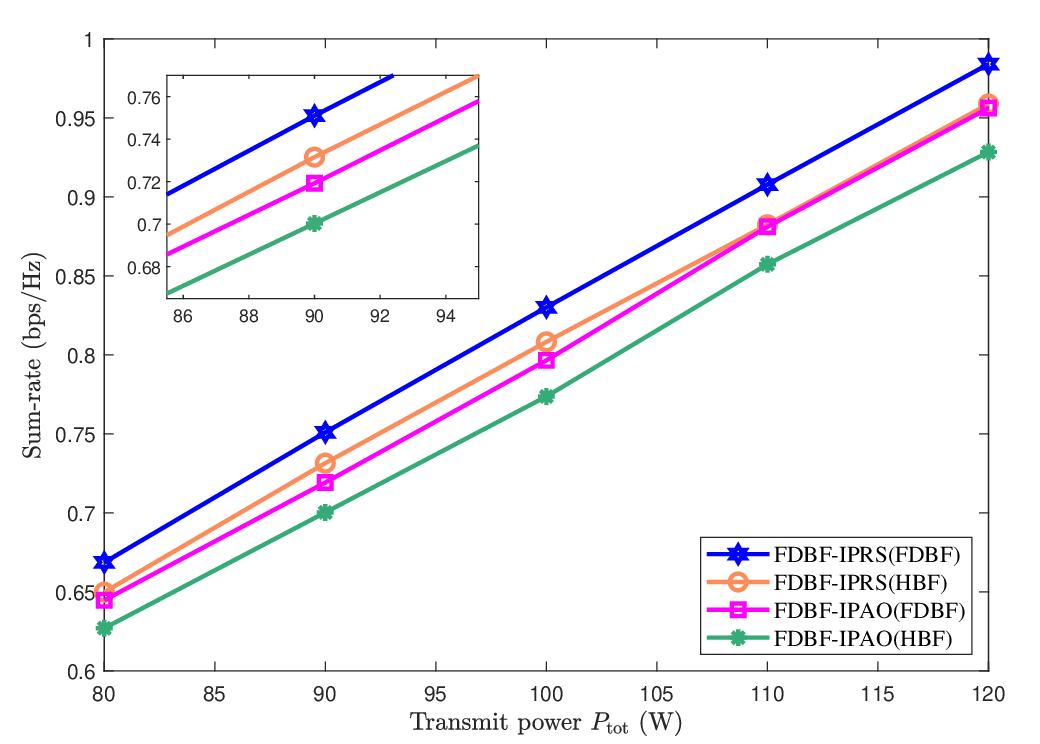}
	\caption{Comparisons of sum-rate of FDBF-IPRS and FDBF-IPAO schemes under different $P_{\rm tot}$.~~~~~~~~~~~~~~~~~~~~~~~~~~~~~~~~~~~~~~~~~~~~~~~~~~~~~~~~~~~~~~~~}
	\label{PtotDiffSchemes.Fig}
\end{figure}

To evaluate the convergence of the FDBF-IPAO scheme, we provide the sum-rate versus the number of iterations for different $P_{\rm {tot}}$, as shown in Fig.~\ref{Iteration.Fig}. We set $\gamma_n = 0.01$~bps/Hz, $N_{\rm BS} = 32$ and $N_{\rm s} = 6$. 
When $P_{\rm tot}$ grows from $80$ to $120$~W, the sum-rate can be increased from 0.65~bps/Hz to 0.97~bps/Hz with 50\% improvement. Moreover, it can be seen that the FDBF-IPAO scheme can converge in no more than eight iterations.

As shown in Fig.~\ref{PtotDiffSchemes.Fig}, we compare the sum-rate under different $P_{\rm {tot}}$ for FDBF and HBF with fixed $\gamma_n = 0.01$~bps/Hz, $N_{\rm BS} = 32$ and $N_{\rm s} = 6$, where different schemes including FDBF-IPRS and FDBF-IPAO are employed. It can be observed that as $P_{\rm {tot}}$ increase, the sum-rate of total $N_{\rm s}$ beam positions increases for both FDBF and HBF. For either FDBF-IPRS or FDBF-IPAO, the sum-rate of HBF is lower than that of FDBF, since HBF has constraints from the phase shifter networks and achieves lower hardware complexity than FDBF. For the same FDBF, the sum-rate achieved by the FDBF-IPAO scheme is slightly lower than that achieved by the FDBF-IPRS scheme. The reason is that the FDBF-IPAO scheme reduces the computational complexity by solving the continuous optimization problem for illumination pattern design, where the random initialization of the illumination pattern may lead to a fast convergence to a locally optimal solution.

\begin{figure}[!t]
	\includegraphics[width=0.5\textwidth]{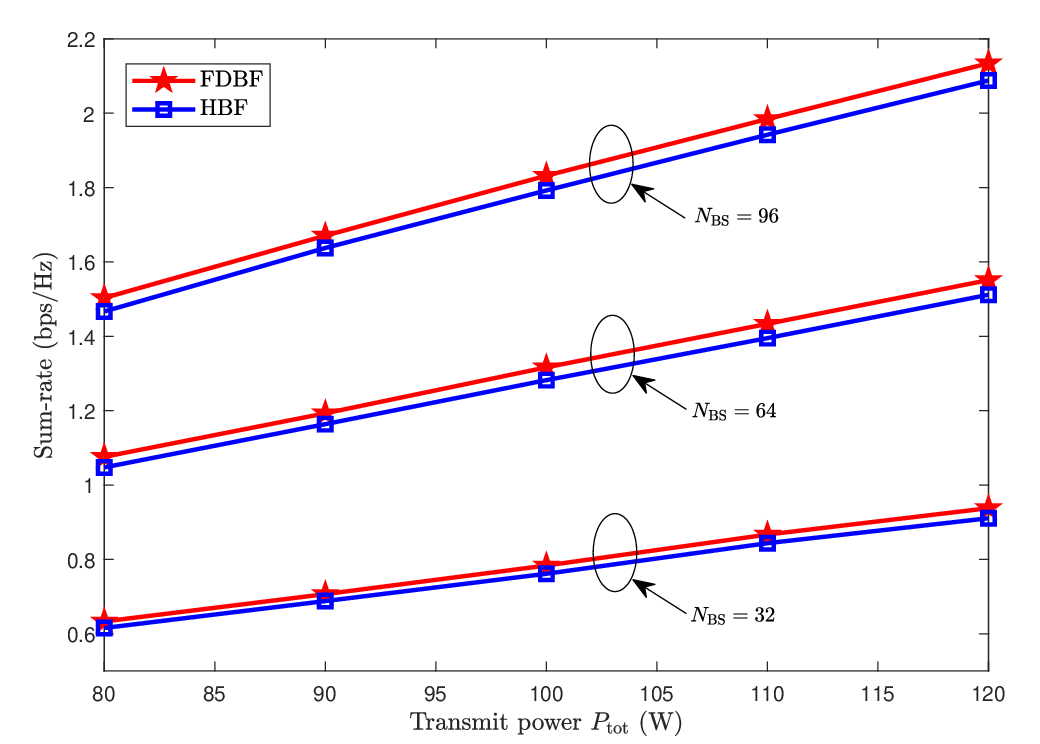}
	\caption{Comparisons of sum-rate of total $N_{\rm s}$ beam positions under different $P_{\rm tot}$ for FDBF and HBF equipped with different $N_{\rm BS}$.~~~~~~~~~~~~~~~~~~~~~~~~~~~~~}
	\label{NbsnumberPtot.Fig}
\end{figure}

\begin{figure}[!t]
	\includegraphics[width=0.5\textwidth]{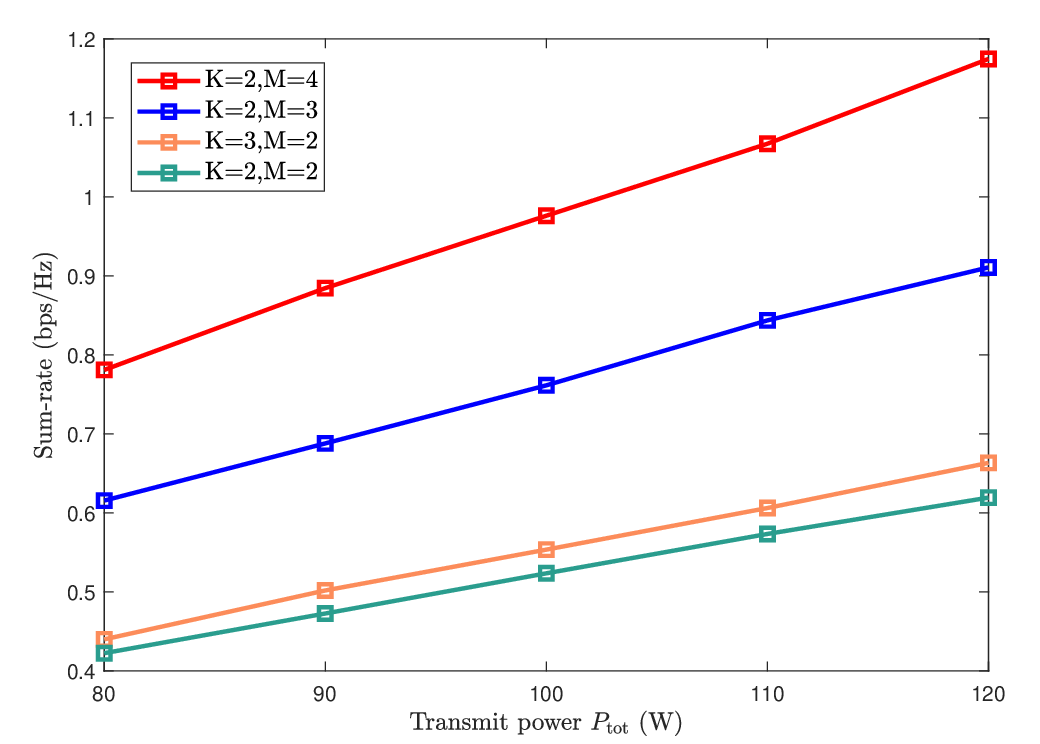}
	\caption{Comparisons of sum-rate of total $N_{\rm s}$ beam positions for different $K$ and $M$. ~~~~~~~~~~~~~~~~~~~~~~~~~~~~~~~~~~~~~~~~~~~~~~~~~~~~~~~~~~~~~~~~~~~~~~~~~~~~~~~~}
	\label{PtotNsnumber.Fig}
\end{figure}

Fig.~\ref{NbsnumberPtot.Fig} compares the sum-rate of total $N_{\rm s}$ beam positions for FDBF and HBF using the FDBF-IPAO scheme, where we set different $P_{\rm tot}$ and $N_{\rm BS}$ and fix $\gamma_n = 0.01$~bps/Hz. It is seen that larger $P_{\rm{tot}}$ leads to higher sum-rate. To further evaluate the impact of different antenna numbers, we set $N_{\rm BS} = 32$, $64$ and $96$ for both FDBF and HBF design schemes. When we enlarge $N_{\rm BS}$ from $32$ to $96$, the increased sum-rate can be achieved for the same $P_{\rm tot}$. Furthermore, as $N_{\rm BS}$ increases, the gap between FDBF and HBF becomes large. The reason is that more antennas lead to more unit-modulus constraints coming from the phase shifters for HBF, which makes HBF more difficult to approach the performance of FDBF. 

As shown in Fig.~\ref{PtotNsnumber.Fig}, we compare the sum-rate for different $K$ and $M$ using the FDBF-IPAO scheme for the HBF design. We fix $\gamma_n = 0.01$~bps/Hz and $N_{\rm BS} = 32$. In fact, we compare $N_{\rm s}=4$, $N_{\rm s}=6$ and $N_{\rm s}=8$. It is seen that as $N_{\rm s}$ increases, the sum-rate substantially improves. When fixing $K=2$, either enlarging $M=2$ to $M=3$ or enlarging $M=3$ to $M=4$ leads to around 40\% increase of the sum-rate, which implies that more time slots for SatComs can effectively improve the sum-rate. When fixing $M=2$, enlarging $K=2$ to $K=3$ leads to around 6\% increase of the sum-rate. When fixing $N_{\rm s}=6$, the sum-rate of $K=2$ and $M=3$ is around 38\% higher than that of $K=3$ and $M=2$, which implies that improving the number of time slots for SatComs is more effective than increasing the number of illuminated beam position within the same time slot.

%The reason is that high sum-rate threshold of each beam position consumes much transmit power and higher sum-rate can be achieved for the same scheme with increased number of beam positions.

%-----------------------------------------------------------------------------------------------------

\section{Conclusion}
In this paper, we have investigated the HBF design for BH LEO SatComs. Aiming at maximizing the sum-rate of totally illuminated beam positions during the whole BH period, we have considered the joint BIP design problem subject to the HBF constraints and sum-rate requirements. To solve this joint BIP design problem, we have temporarily replaced the HBF constraints with the FDBF constraints and converted the original problem into a fully-digital BIP design problem. We have proposed the FDBF-IPRS and FDBF-IPAO schemes to design the fully-digital beamformers. Then we have proposed the HBF-AM algorithm to design the hybrid beamformers. Simulation results have shown that the proposed schemes can achieve satisfactory sum-rate performance for BH LEO SatComs. Future work will be continued with the focus on the power allocation for BH LEO SatComs.

%\section*{Acknowledgment}
%This work is supported in part by National Key Research and Development Program of China under Grant 2021YFB2900404. 

%\bibliographystyle{IEEE-unsorted}
\bibliographystyle{IEEEtran}
\bibliography{refs.bib}

\end{document}